\documentstyle[11pt,newpasp,twoside,epsf]{article}

\def\edcomment#1{\iffalse\marginpar{\raggedright\sl#1\/}\else\relax\fi}
\reversemarginpar

\begin{document}
\title{Multi-wavelength observations of the red giant R~Doradus with the
MAPPIT interferometer}
\author{A.P.~Jacob$^1$, T.R.~Bedding$^1$, J.G.~Robertson$^1$,
J.R.~Barton, C.A.~Haniff, R.G.~Marson \& M.~Scholz}
\affil{$^1$School of Physics, University of Sydney 2006, Australia}

\begin{abstract}
We present visibility measurements of the nearby Mira-like star R~Doradus
taken over a wide range of wavelengths (650--990\,nm).  The observations
were made using MAPPIT (Masked APerture-Plane Interference Telescope), an
interferometer operating at the 3.9-m Anglo-Australian Telescope.  We used
a slit to mask the telescope aperture and prism to disperse the
interference pattern in wavelength.  We observed in R~Dor strong decreases
in visibility within the TiO absorption bands.  The results are in general
agreement with theory but differ in detail, suggesting that further work is
needed to refine the theoretical models.
\end{abstract}

Improvements in atmospheric models for M giants requires measurements of
stellar radii over a wide range of wavelengths.  We present such
observations for R~Doradus (HR~1492; $V=5.4$; spectral type M8\,IIIe), a
nearby Mira-like star.  We used MAPPIT (Masked APerture-Plane Interference
Telescope), an interferometer operating at the 3.9-m Anglo-Australian
Telescope.  The telescope aperture was masked with a slit and a prism was
used to disperse the interference pattern in wavelength.

\begin{figure}
\plotone{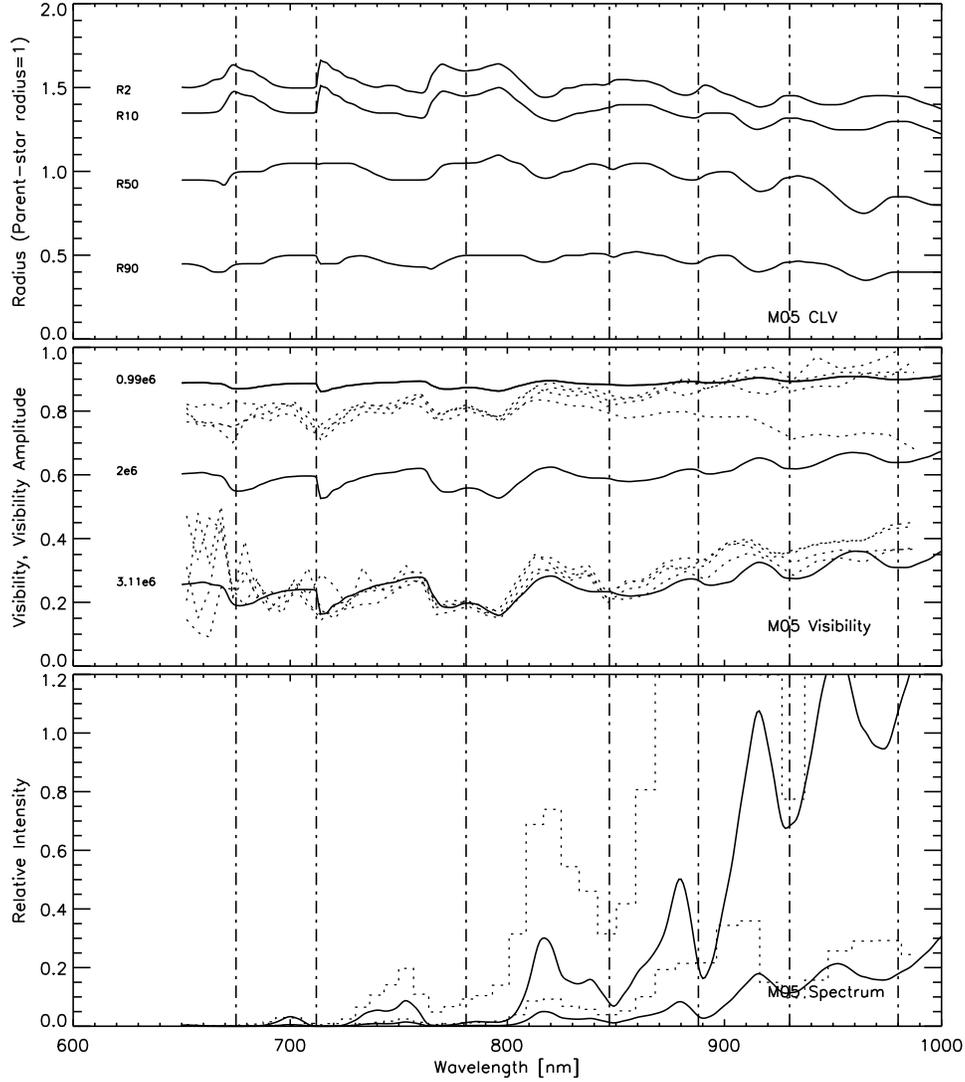}		
\caption[]{\label{p84ihere} Comparison of observations of R~Dor (dotted
curves) with a 1.2M$_{\odot}$ fundamental-mode model (M96400).  The
Rosseland angular diameter of the model was set to 53\,mas.  The top panel
shows the model radius at 2, 10, 50 and 90\,\% of the central intensity.
The middle panel shows model visibilities at three spatial frequencies, and
the observed visibilities at two spatial frequencies.  The bottom panel
shows the flux spectrum for the model and observations, both plotted twice
with vertical scales differing by a factor of 20 to accommodate the large
range in intensities.  To guide the eye, vertical dashed lines indicate the
strongest absorption bands.  The model calculations are described in full
by Jacob et al.\ (2000a) and further details of the observations and
discussion of the model fits are given by Jacob et al.\ (2000b).  }
\end{figure}

We measured visibility over a wide range of wavelengths and compared our
observations with the M-giant models by Bessell, Brett, Hofmann, Scholz and
Wood (see Hofmann et~al.\ 1998 and references within).  We find agreement
in the general features, with clear drops in visibility within the TiO
absorption bands.  However, although some models match the visibility
observations in some regions, none match across the whole spectrum.  The
static 1M$_{\odot}$ models, in particular, did not fit our observations
well.  The 1.2M$_{\odot}$ M-series models (and 1M$_{\odot}$ D models), both
of which pulsate in the fundamental mode, appear to be the best fits to our
observations (see Fig.~\ref{p84ihere}).


\end{document}